 \documentclass[final,5p,times,twocolumn]{elsarticle}

\usepackage{amssymb}
\usepackage{amsmath}
\usepackage{amsfonts}
\usepackage{algorithmic}
\usepackage{algorithm}
\usepackage{array}
\usepackage[caption=false,font=normalsize,labelfont=sf,textfont=sf]{subfig}
\usepackage{textcomp}
\usepackage{stfloats}
\usepackage{url}
\usepackage{verbatim}
\usepackage{graphicx}
\usepackage{arydshln}
\usepackage{bm}
\usepackage{color}

\journal{Elsevier}

\newcommand{\br}{{\mathbb R}}
\newcommand{\bz}{{\mathbb Z}}

\newcommand{\bc}{{\mathbb C}}

\newcommand{\spann}{{\rm span}}

\newcommand{\diag}{{\rm diag}}

\newcommand{\prf}{\par{\bf Proof. }}
\newcommand{\bbox}{\rule{2mm}{2.5mm}}

\newcommand{\bfone}{\mathbf{1}}

\newcommand{\bfx}{\mathbf{x}}

\newcommand{\bfa}{\mathbf{a}}
\newcommand{\bfb}{\mathbf{b}}
\newcommand{\bfc}{\mathbf{c}}
\newcommand{\bfu}{\mathbf{u}}
\newcommand{\bfU}{\mathbf{U}}
\newcommand{\bfS}{\mathbf{S}}
\newcommand{\bfH}{\mathbf{H}}
\newcommand{\bfL}{\mathbf{L}}

\newtheorem{thm}{Theorem}[section]
\newtheorem{lem}[thm]{Lemma}
\newtheorem{de}{Definition}
\newtheorem{coro}[thm]{Corollary}
\newtheorem{prop}[thm]{Proposition}

\begin{document}

\begin{frontmatter}



\title{Atomic Filter: a Weak Form of Shift Operator for Graph Signals}

\author[1,2]{Lihua Yang}
\ead{mcsylh@mail.sysu.edu.cn}
\author[1]{Qing Zhang}
\ead{zhangq369@mail2.sysu.edu.cn}
\author[3,4]{Qian Zhang}
\ead{mazhangq@szu.edu.cn}
\author[3,4]{Chao Huang\corref{cor1}}
\ead{hchao@szu.edu.cn}

\cortext[cor1]{Corresponding author}

\address[1]{School of Mathematics, Sun Yat-sen University, Guangzhou 510275, China}
\address[2]{Guangdong Province Key Laboratory of Computational Science}
\address[3]{College of Mathematics and Statistics, Shenzhen University, Shenzhen 518060, China}
\address[4]{Shenzhen Key Laboratory of Advanced Machine Learning and Applications, Shenzhen University, Shenzhen 518060, China}




\begin{abstract}

The shift operation plays a crucial role in the classical signal processing.
It is the generator of all the filters and the  basic operation for time-frequency analysis,
such as windowed Fourier transform and wavelet transform.
With the rapid development of internet technology and big data science, a large amount of data are expressed as signals defined on graphs.
In order to establish the theory of filtering, windowed Fourier transform and wavelet transform  in the setting of graph signals, we need to extend the shift operation of classical signals to graph signals.
\par
It is a fundamental problem since the vertex set of a graph is usually not a vector space and the addition operation cannot be defined on the vertex set of the graph.
In this paper, based on our understanding on the core role of shift operation in classical signal processing
we propose the concept of atomic filters, which can be viewed as a weak form of the shift operator for graph signals. Then, we study the conditions such that an
atomic filter is norm-preserving, periodic, or real-preserving.
The property of real-preserving holds naturally in the classical signal processing, but
no the research has been reported on this topic in the graph signal setting.
With these conditions
we propose the concept of normal atomic  filters for graph signals, which degenerates
into the classical shift operator under mild conditions if the graph is circulant.
Typical examples of graphs that have or have not
normal atomic filters are given.
Finally, as an application, atomic filters are utilized to construct time-frequency atoms
which constitute a frame of the graph signal space.
\end{abstract}




\end{frontmatter}


\section{Introduction}
Graphs provide a natural representation for data in many applications, such as social
networks, web information analysis, sensor networks and machine learning \cite{Shuman-2013,Sandryhaila-2014,Ortega-2018}.
Graph signals are functions defined on the vertices of graphs. To process such signals,
one needs to extend the well-developed theory of classical signal processing to graph signals.
There have been a lot of researches on graph signal processing, including
graph shift operators \cite{Shuman-2016,Sandryhaila-2013,Girault-2015,Girault-bk2015,Girault-2016,Gavili-2017,Grelier-2016,Pasdeloup-2018,Dees-2019,huang2021approximation},
graph filters \cite{Sandryhaila-2014b,Segarra-2015,Tremblay-2016,Chen-2018,Sakiyama-2019},
graph Fourier transforms \cite{Sandryhaila-2013b,Sardellitti-2017,Yang},
windowed graph Fourier transforms \cite{Shuman-2016,Shuman-2012},
graph wavelets \cite{Coifman-2006,Hammond-2011,Leonardi-2013,Dong-2017,Nakahira-2018},
graph signal sampling \cite{Pesenson-2008,Chen-2015,Wang-2015,Anis-2016,Marques-2016,Puy-2018,Chamon-2018,Huang-2020,Yang-2021a,Yang-2021b},
multiscale analysis \cite{Gavish-2010,Shuman-2016b,Avena-2020},
and approximation theory for graph signals \cite{Pesenson-2010b,Pesenson-2010,Zhu-2012}.
\par
The shift operation play a crucial role in the classical signal processing. It is the generator
of filters and the basic operation of defining windowed Fourier transform
and wavelet transform \cite{oppenheim1999discrete,mallat1999wavelet}.
It is also used to define the moduli of smoothness of functions \cite{DeVore-bk1993}.
However, the definition of the classical shift operator $Sx(t):=x(t-h)$ cannot be
formally extended  to the graph
signal setting since the vertex set of a graph is usually not a vector space, which makes it impossible to perform addition operation $x+h$ between the vertices of the graph.
It is a fundamental problem of graph signal processing, both in theory and application.
In recent years, from different perspectives of classical shift operation,
several different definitions for graph shift operators have been proposed.
D. I. Shuman \textit{et al.} presented a definition via the generalized
convolution with a delta function centered at a given vertex \cite{Shuman-2016}.
A. Sandryhaila \textit{et al.} utilized the adjacency matrix of the graph as the shift operator \cite{Sandryhaila-2013}.
In \cite{Girault-2015,Girault-bk2015,Girault-2016}, B. Girault \textit{et al.} defined a norm-preserving graph shift operator
depending on the spectrum of the graph Laplacian matrix.
Later in \cite{Gavili-2017}, A. Gavili \textit{et al.} defined a set of norm-preserving graph shift operators
flexible to accommodate desired properties.
In \cite{Grelier-2016,Pasdeloup-2018}, N. Grelier \textit{et al.} proposed a definition
relying on neighborhood preserving properties.
In \cite{Dees-2019}, B. S. Dees \textit{et al.} employed the maximum entropy principle to
define a general class of shift operators for random signals on a graph.
To define the modulus of smoothness of graph functions, I. Z. Pesenson \textit{et al.}
defined a family of shift operators by using the Schr\"{o}dinger's semigroup of operators generated by the graph
Laplacian $L$ in \cite{Pesenson-2010b}. Inspired by this idea of this definition, a similar  definition was proposed by C. Huang  \textit{et al.} in \cite{huang2021approximation}.
\par
Understanding what role the shift operation plays in signal processing and what basic
properties it must have will help us to establish the theory of shift for graph signals,
which will be essential rather than formal.
In this paper, we systemically examine the role of shift operation in the classical signal processing and summarize the following basic properties of the shift operator:
\par
(1) It is a permutation of the original signal;
\par
(2) It is norm-preserving;
\par
(3) It is smoothness-perserving;
\par
(4) It is periodic;
\par
(5) It is real-preserving;
\par
(6) It is a filter;
\par
(7) Any filter can be expressed as a polynomial of the shift operator.
\par
(8) It is time-invariant.
\par
On this basis, we propose the concept of atomic filters in the setting of graph signals,
which can be viewed as a weak form of the shift operator for graph signals. Then we
study the characterization and implications of the above properties (1)--(8), and
introduce the concept of the so called normal atomic filters. It is shown that a
normal atomic filter degenerates into the classical shift operator under mild conditions if
the graph is circulant. Since not all graphs have normal atomic filters, typical examples
of graphs that have or have not normal atomic filters are discussed.
Finally, as an application, atomic filters are utilized to construct time-frequency atoms
which constitute a frame of the graph signal space.
\par
Throughout this paper we use the following notations and terminologies:
Matrices and vectors are represented by uppercase and lowercase boldface letters, respectively.
The entries of a matrix $\mathbf{A}$ are denoted as $a_{ij}$,
and the entries of a vector $\mathbf{x}$ are denoted as $\mathbf{x}(n)$ or $x_n$.
Transpose and Hermitian (conjugate transpose) operations are represented by $(\cdot)^T$ and $(\cdot)^*$, respectively.
For a vector $\bfa\in\bc^N$,
the notation $\diag(\mathbf{a})$ represents a diagonal matrix with entries of the vector $\mathbf{a}$ along the main diagonal.
For $\mathbf{x},\mathbf{y}\in\bc^N$,
their inner product is defined as $\langle\mathbf{x},\mathbf{y}\rangle:=\mathbf{y}^*\mathbf{x}$,
and $\|\mathbf{x}\|_2:=\sqrt{\mathbf{x}^*\mathbf{x}}$ is the Euclidean norm of $\mathbf{x}$.
For positive integers $m,n$, the notation $\delta_{m,n}$ is defined as $\delta_{m,n}=1$ for $m=n$ and $\delta_{m,n}=0$ for $m\neq n$.

\section{The Role of Shift Operator in the Classical Signal Processing}
\label{sec:2}
\subsection{Shift Operator in the Classical Signal Processing}
Let us first briefly review the
concepts of shift operators in the classical signal processing.
Without losing generality, we consider the discrete periodic signals.
A discrete signal $\{x_n\}_{n\in\bz}$ is called periodic with period $N$ if
$x_{n+N}=x_n$ holds for any $n\in\bz$. Then the shift operator $\mathbf{S}$ is defined as
\begin{equation}\label{eq:classical-shift}
	(\mathbf{S}x)_n:=x_{n-1},~~~~n\in\bz.
\end{equation}
It is easy to see that this signal can be expressed as an $N$-dimensional column vector
$\mathbf{x}:=(x_1,\cdots, x_N)^T$ and the shift operator $\mathbf{S}$ can be expressed as
the following circulant matrix:
\begin{equation}\label{eq:shift-S}
	\mathbf{S}=
	[e_2,...,e_N, e_1],
\end{equation}
where $I_N=[e_1,...,e_N]$ is the identity matrix of order $N$.
For notational simplicity, we shall use in this paper the same notation
$\bfx$ for the signal $\{x_n\}_{n\in\bz}$ and the vector $\bfx$;
the notation $\mathbf{S}$ for the shift operator \eqref{eq:classical-shift} and the matrix
\eqref{eq:shift-S}: the context should make the distinction clear.
\par
According to \eqref{eq:classical-shift} or \eqref{eq:shift-S}, it is easy to see that
$\mathbf{S}$ is a downshift permutation operator which pushes the components of a vector
down one notch with wraparound. Hence, it possesses the following basic properties:
\par
(1) $\mathbf{S}\bfx$ is a permutation of $\bfx$, as described above.
\par
(2) Norm-preserving property: $\|\bfS\bfx\|_2=\|\bfx\|_2$;
\par
(3) Smoothness-perserving: $\sigma(\mathbf{S}\bfx)=\sigma(\bfx)$ holds for
$$\sigma(\bfx):=\sum^N_{i=1}(|x_i-x_{i-1}|^2+|x_i-x_{i+1}|^2).$$
\par
(4) Periodicity: $\mathbf{S}^N\mathbf{x}=\mathbf{x}$;
\par
(5) Real-preserving property: If $\mathbf{x}$ is real-valued, then
$\mathbf{S}\mathbf{x}$ is also real-valued.
\par
Similarly, we have 2-dimensional shift operators:
\begin{equation}\label{eq:classical-shift-2D}
	\begin{cases}
		(\mathbf{S}_1x)_{n,m}:=x_{n-1,m},\\
		(\mathbf{S}_2x)_{n,m}:=x_{n,m-1},\\
		(\mathbf{S}_3x)_{n,m}:=x_{n-1,m-1},
	\end{cases}~~~~n, m\in\bz.
\end{equation}
It is easy to verify that the three shifts satisfy the above conditions (1)--(5) for
\begin{align*}
	\sigma(\bfx):=&\sum^N_{i,j=1}(|x_{i,j}-x_{i-1,j}|^2+|x_{i,j}-x_{i+1,j}|^2\\
	&~~+|x_{i,j}-x_{i,j-1}|^2+|x_{i,j}-x_{i,j+1}|^2).
\end{align*}
\par
The above properties or Definitions \eqref{eq:classical-shift} and
\eqref{eq:classical-shift-2D} are not convenient to be
used to extend the shift operators to graph signals.
As discussed in the previous section, to extend the concept of shift operator
to graph signals, we need to examine systemically the crucial role of the shift operator
in the classical signal processing.

\subsection{As a Generator of Filters}
In the classical signal processing, the discrete Fourier transform and the inverse discrete Fourier transform of $\mathbf{x}$ are respectively defined as
$$\hat{\mathbf{x}}:=\mathbf{U}^{-1}\mathbf{x},~~~~\mathbf{x}=\mathbf{U}\hat{\mathbf{x}},$$
where $\mathbf{U}=(\bfu_1,\cdots,\bfu_N)$ is the matrix of discrete Fourier basis, whose $k$-th column is given by
\begin{equation}\label{eq-FT}
	\bfu_k :=\frac{1}{\sqrt{N}}(1, \omega^{k-1}, \omega^{2(k-1)}\cdots,\omega^{(N-1)(k-1)})^T,
\end{equation}
where $\omega:=\exp(2\pi i/N)$.
\par
Given a vector $\mathbf{h}:=(h_0,\cdots,h_{N-1})^T\in\bc^N$, a filter corresponding to
$\mathbf{h}$ is defined by
\begin{equation}\label{eq:FT-classical}
	(\widehat{\mathbf{H}\mathbf{x}})(k)
	:=\hat{\mathbf{x}}(k)\,\hat{\mathbf{h}}(k),~~~~k=1,\cdots,N,
\end{equation}
i.e.,
$$(\mathbf{H}\mathbf{x})(n)=\frac{1}{\sqrt{N}}\sum_{k=0}^{N-1}h_kx_{n-k}
=\frac{1}{\sqrt{N}}\sum_{k=0}^{N-1}h_k(\mathbf{S}^k\mathbf{x})(n),$$\\
$n=1,\cdots,N,$
or equivalently,
\begin{equation}\label{eq:filtering-1}
	\mathbf{H}=\frac{1}{\sqrt{N}}\sum_{k=0}^{N-1}h_k\mathbf{S}^k,
\end{equation}
where $\mathbf{S}$ is the shift operator defined by \eqref{eq:shift-S}.
Eq.~\eqref{eq:filtering-1} shows
that the shift operator is a generator of any filters in the sense that
\par
(6) The shift operator is a filter (corresponding to
$\mathbf{h}:=\sqrt{N}(0,1,0,\cdots, 0)^T$);
\par
(7) Any filter can be expressed as a polynomial of the shift operator.
\par
Besides, Eq.~\eqref{eq:filtering-1} implies that the composition of $\mathbf{H}$ and $\mathbf{S}$ are commutative:
\begin{equation}\label{eq:commutativity}
	\mathbf{H}\mathbf{S}=\mathbf{S}\mathbf{H}.
\end{equation}
This equality is called the time-invariance of $\mathbf{H}$. It means that
$\mathbf{H}(\bfx(t-t_0))=(\mathbf{H}\bfx)(t-t_0)$,  in other words,
the time-shift by $t_0$ of the input signal creates the same time-shift by $t_0$ at
the output. As proved in \cite[\S 2.1]{mallat1999wavelet}, the time-invariance is also
a sufficient condition for a linear operator to be a filter.
Thus, the shift operator serves as a descriptor of filters in the following sense:
\par
(8) A linear operator $\mathbf{H}$ is a filter if and only if it is commutative with
$\mathbf{S}$, i.e., \eqref{eq:commutativity} holds.

\section{Atomic Filters for Graph Signals}
\subsection{Graph Fourier Transforms and Graph Filters}
\label{sec:GSP}
\par
In this paper, we consider connected, weighted, and undirected graphs.\footnote{
	Most of definitions and results in this paper can be extended to a directed graph
	provided an orthonormal graph Fourier basis is available.
}
Let $\mathcal{V}=\{v_1,\cdots,v_N\}$ be
the set of vertices of a graph $\mathcal{G}$, and $\mathbf{W}\in\br^{N\times N}$ be the weighted adjacency matrix
with its entry $w_{ij}$ the nonnegative weight of the edge between the vertices $v_i$ and $v_j$.
A graph signal is a function defined on $\mathcal{V}$ and
can be expressed as a vector $\bfx\in\bc^N$,
whose $n$-th component represents the function value at the $n$-th vertex.
\par
The Laplacian matrix is defined by $\mathbf{L}:=\mathbf{D}-\mathbf{W}$,
where $\mathbf{D}$ is the degree matrix $\diag(d_1,\cdots,d_N)$ with $d_i:=\sum_{j=1}^Nw_{ij}$. By the spectral decomposition we have that
$\mathbf{L}=\mathbf{U}\bm{\Lambda}\mathbf{U}^{-1}$, where
$$\mathbf{U}:=(\mathbf{u}_1,\cdots,\mathbf{u}_N),~~~~
\bm{\Lambda}:=\diag(\lambda_1,\cdots,\lambda_N),$$
with $0=\lambda_1<\lambda_2\le\cdots\le\lambda_N$ being all the eigenvalues of
$\mathbf{L}$ and $\mathbf{u}_1,\cdots,\mathbf{u}_N$ being a set of
eigenvectors, which constitute an orthonormal basis of $\bc^N$ and
$\mathbf{u}_1=\bfone/\sqrt{N}$, where $\bfone\in\br^N$ is an $N$-dimensional vector of ones.
To accommodate desired properties for graph filters,
we allow the eigenvectors $\mathbf{u}_2,\cdots,\mathbf{u}_N$ to be complex vectors.
\par
For a graph signal $\bfx\in\bc^N$, its graph Fourier transform and
inverse graph Fourier transform are respectively defined as
$$\hat{\mathbf{x}}:=\mathbf{U}^{-1}\mathbf{x},~~~~
\mathbf{x}=\mathbf{U}\hat{\mathbf{x}}.$$
Hereafter, we call $\{\mathbf{u}_k\}_{k=1}^N$ the graph Fourier basis and $\mathbf{U}$ the matrix of graph Fourier basis.
\par
Using the graph Fourier transform, the graph filter can be defined in a way similar to Eq.~\eqref{eq:FT-classical}.
Given a vector $\mathbf{h}\in\bc^N$, the corresponding graph filter $\mathbf{H}$ is defined as a linear
operator satisfying \cite{Shuman-2013,Girault-2015}:
\begin{equation}\label{eq:de:graph-filter}
	(\widehat{\mathbf{H}\mathbf{x}})(k)
	:=\hat{\mathbf{x}}(k)\,\hat{\mathbf{h}}(k),~~~~k=1,\cdots,N,
\end{equation}
for any graph signal $\mathbf{x}\in\bc^N$.
\par
It is easy to see that Eq.~\eqref{eq:de:graph-filter} holds for any $\mathbf{x}\in\bc^N$
if and only if
$\mathbf{H}=\mathbf{U}\diag(\hat{\mathbf{h}})\mathbf{U}^{-1}$.
Thus, any graph filter can be expressed as the following form
\begin{equation}\label{eq:graph-filter}
	\mathbf{H}_\mathbf{a}:=\mathbf{U}\diag(\mathbf{a})\mathbf{U}^{-1},
\end{equation}
where the vector $\mathbf{a}\in\bc^N$ is called the frequency response of the filter
\cite{Sandryhaila-2014,Sandryhaila-2013,Sandryhaila-2014b}.

\subsection{Definition of Atomic Filters for Graph Signals}
As discussed in Section \ref{sec:2}, the classical shift operator $\mathbf{S}$ satisfies
the properties (1)--(8). It will be shown that such operator that satisfies all the
eight properties does not always exist for every graphs.
Therefore, in order to extend the classical shift operator to graph signals, in this paper, we do not
try to establish the shift operators with property (1)--(8), but use Property (6) and (7) to
introduce the concept of ``atomic filters'' for general graphs, which can be viewed as a weak form of
the shift operator for graph signals. Then we shall discuss the existence and
conditions for atomic filters to satisfy the properties (1)--(8).
\begin{de}\label{de:atom-filter}
	A matrix $\mathbf{S}\in\bc^{N\times N}$ is called an atomic filter if
	any graph filter can be expressed as a polynomial of $\mathbf{S}$.
\end{de}
\par
We first point out that an atomic filter is a filter, i.e. (6) holds. In fact, suppose
any graph filter $\mathbf{H}_\mathbf{b}=\mathbf{U}\diag(\mathbf{b})\mathbf{U}^{-1}$
with $\mathbf{b}\in\bc^N$ can be expressed as a polynomial of $\mathbf{S}$,
then $\mathbf{H}_\mathbf{b}\mathbf{S}=\mathbf{S}\mathbf{H}_\mathbf{b}$, i.e.,
$$\diag(\mathbf{b})\mathbf{A}=\mathbf{A}\diag(\mathbf{b})$$
for $\mathbf{A}:=(a_{ij}):=\mathbf{U}^{-1}\mathbf{S}\mathbf{U}$.
Consider the $(i,j)$-th element on both sides of the above equation, we get
$$b_ia_{ij}=a_{ij}b_j,~~~~i,j=1,\cdots,N.$$
Taking a vector $\mathbf{b}\in\bc^N$ with distinct components,
we deduce that $a_{ij}=0$ for any $i\neq j$, which implies that
$\mathbf{A}=\diag(\mathbf{a})$ for some vector $\mathbf{a}\in\bc^N$ and consequently
$$\mathbf{S}=\mathbf{U}\mathbf{A}\mathbf{U}^{-1}
=\mathbf{U}\diag(\mathbf{a})\mathbf{U}^{-1}=\mathbf{H}_\mathbf{a}$$
is a filter.
\par
It is a natural question to investigate the construction of atomic filters, that is,
to make clear what $\mathbf{a}\in\bc^N$ makes $\mathbf{H}_\mathbf{a}$ satisfy
the condition of Definition \ref{de:atom-filter}. The following theorem gives a
sufficient and necessary condition to this question.
\begin{thm}\label{th:ch1-shift-def}
	A graph filter $\mathbf{H}_\mathbf{a}$ is an atomic filter if and only if
	the vector $\mathbf{a}\in\bc^N$ has distinct components.
\end{thm}
\prf
The filter $\mathbf{H}_\mathbf{a}$ is an atomic filter if and only if any graph filter
$\mathbf{H}_{\mathbf{b}}=\mathbf{U}\diag(\mathbf{b})\mathbf{U}^{-1}$
with $\mathbf{b}\in\bc^N$ can be expressed as
\begin{equation}\label{eq:pre-shift-condition-0}
	\mathbf{H}_\mathbf{b}=\sum^{N-1}_{k=0}c_k\mathbf{H}_\mathbf{a}^k
\end{equation}
for some constants $c_0,c_1,\cdots, c_{N-1}\in\bc$. Here, in the right hand of
Eq.~\eqref{eq:pre-shift-condition-0}, the highest degree of the polynomial is $N-1$
due to the Cayley-Hamilton theorem \cite[pp.~109]{Horn-bk2013}.
It is easy to verify that Eq.~\eqref{eq:pre-shift-condition-0} is equivalent to
$$\diag(\mathbf{b})=\sum^{N-1}_{k=0}c_k[\diag(\mathbf{a})]^k,$$
i.e.,
$$b_n = \sum^{N-1}_{k=0}c_ka_n ^k,~~~~n =1,\cdots,N,$$
which is further equivalent to $\mathbf{A}\mathbf{c}=\mathbf{b}$ for the
Vandermonde matrix $\mathbf{A}$ determined by $a_1,...,a_N$.
Hence, $\mathbf{H}_\mathbf{a}$ is an atomic filter if and only if the matrix
$\mathbf{A}$ is invertible, namely, the vector $\mathbf{a}\in\bc^N$ has distinct components.
\bbox
\par
\textbf{Remark}: Theorem \ref{th:ch1-shift-def} shows that an atomic filter
$\mathbf{S}=\bfH_\bfa$ is a nonderogatory matrix. According to
\cite[p.~178]{Horn-bk2013}, a linear operator $\mathbf{H}$ is commutative
with $\mathbf{S}$ if and only if $\mathbf{H}$ is a
polynomial of $\mathbf{S}$. This means  an atomic filter satisfies the property (8).

\subsection{Basic Properties of Atomic Filters}
It has been already known that an atomic filter $\mathbf{H}_\bfa$ must satisfies the properties
(6)--(8).
In this section, we investigate whether or under what conditions an atomic filter satisfies
the properties (1)--(5). Property (1) is obviously true in the classical signal processing. In graph signal processing,
it is equivalent to the existence of a permutation matrix $\mathbf{P}$ such that
$\mathbf{U}\diag(\bfa)\mathbf{U}^{-1}=\mathbf{P}$. It will be shown that this
condition is usually not true in graph settings. Below, let us study Properties (2)--(5).
\par
An atomic filter $\bfH_a$ is called norm-preserving if $\|\bfH_\bfa\bfx\|_2=\|\bfx\|_2$
holds for any $\bfx\in\bc^N$.  It is smoothness-perserving if
$\sigma(\mathbf{S}\bfx)=\sigma(\bfx)$ holds for
$$\sigma(\bfx):=\bfx^*\mathbf{L}\bfx
=\sum^N_{k=1}\lambda_k|\hat{\bfx}(k)|^2,$$
where $\mathbf{L}$ is the Laplacian matrix and $0=\lambda_1\le\cdots\le\lambda_N$
are all the eigenvalues of $\mathbf{L}$.
\begin{prop}\label{th:isometricity}
	Let $\mathbf{H}_\mathbf{a}$ be an atomic filter. Then
	\par
	(1) It  is norm-preserving if and only if $|a_k|=1$ for $k=1,\cdots,N$.
	\par (2) If it is norm-preserving, then it is smoothness-preserving.
	\par
	(3) It is periodic, i.e., $\mathbf{H}_\mathbf{a}^N=\mathbf{I}$, if and only if $a_k=e^{-i\theta_k},~k=1,\cdots,N$,
	where
	\begin{equation}\label{eq:period}
		\{\theta_k\,|\,k=1,\cdots,N\}=\Big\{\frac{2\pi}{N}(k-1)\,\Big|\,k=1,\cdots,N\Big\}.
	\end{equation}
\end{prop}
\prf
(1) By
\begin{align*}
	&\|\diag(\mathbf{a})\bfx\|_2=\|\bfU\diag(\mathbf{a})\bfU^{-1}\bfU\bfx\|_2
	=\|\mathbf{H}_\mathbf{a}\bfU\bfx\|_2,\\
	&\|\bfU\bfx\|_2=\|\bfx\|_2,
\end{align*}
we have that $\bfH_\bfa$ is norm-preserving if and only if
$\|\mathbf{H}_\mathbf{a}\bfU\bfx\|_2=\|\bfU\bfx\|_2$, i.e.,
$$\|\diag(\mathbf{a})\bfx\|_2=\|\bfx\|_2,~~~~\forall\bfx\in\bc^N.$$
It is easy to see that this equality is equivalent to $|a_k|=1$ for $k=1,\cdots,N$.
\par
(2) If $|a_k|=1$ for $k=1,\cdots,N$, then for any $\bfx\in\bc^N$ there holds
$$\widehat{\bfH_\bfa\bfx}=\bfU^{-1}\bfH_\bfa\bfx=\bfU^{-1}\bfU\diag(\bfa)\bfU^{-1}\bfx
=\diag(\bfa)\hat{\bfx},$$
which implies $|(\widehat{\bfH_\bfa\bfx})(k)|=|a_k\hat{\bfx}(k)|=|\hat{\bfx}(k)|,~
k=1,\cdots,N$, and consequently $\sigma(\bfH_\bfa\bfx)=\sigma(\bfx)$.
\par
(3) It is easy to see that $\mathbf{H}_\mathbf{a}^N=\mathbf{I}$ is equivalent to
$\mathbf{U}[\diag(\mathbf{a})]^N\mathbf{U}^{-1}=\mathbf{I}$,
which implies $a_k^N=1$, $k=1,\cdots,N$.
\par
Necessity: Let $\mathbf{H}_\mathbf{a}^N=\mathbf{I}$.
Then for $k=1,\cdots,N$, we have $a_k^N=1$, which implies $a_k:=e^{-i\theta_k}$
with $0\le \theta_k<2\pi$. Since $\mathbf{H}_\mathbf{a}$ is an atomic filter,
$\{\theta_k\}_{k=1}^N$ are distinct numbers.
By $a_k^N=1$ we have $\frac{\theta_k}{2\pi}N\in\bz$. Thus
$\{\frac{\theta_k}{2\pi}N\}_{k=1}^N$ are $N$ distinct integers in $[0,N)$.
Since there are only $N$ distinct integers $\{0,1,\cdots,N-1\}$ in $[0,N)$,
we conclude that Eq.~\eqref{eq:period} holds.
\par
Sufficiency: If Eq.~\eqref{eq:period} is satisfied, then $a_k^N=1$ for $k=1,\cdots,N$,
which is equivalent to  $\mathbf{H}_\mathbf{a}^N=\mathbf{I}$.
\bbox
\par
Let us consider the Real-Preserving Property (5).
In the classical signal processing, it is known that the shift operator transforms a
real-valued signal into a real-valued signal. A natural question is whether or under
what conditions this property holds in the graph signal setting. There is no report
on the research on this topic.
\begin{figure*}[!t]
	\centerline{\includegraphics[width=1.0\textwidth]{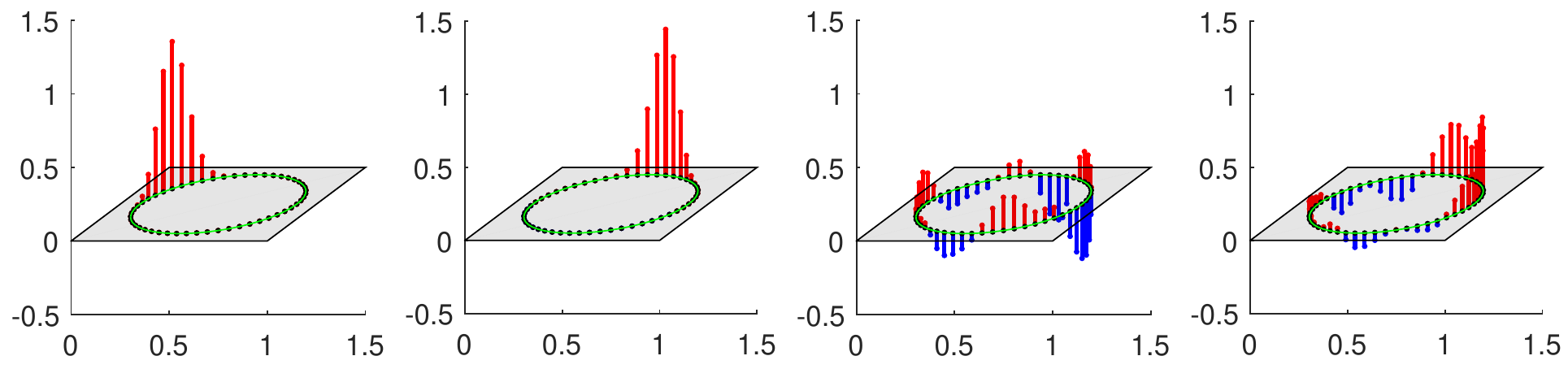}}
	\centerline{\small \hspace{0.03\textwidth} (a)\hspace{0.21\textwidth}
		(b)\hspace{0.21\textwidth}(c)\hspace{0.21\textwidth}(d)}
	\caption{\small (a) A Gaussian signal $\bfx$ defined on a ring graph;
		(b) $\bfH_\bfa^{10}\bfx$; (c) The real part of $\bfH_\bfb^{10}\bfx$;
		(d) The imaginary part of $\bfH_\bfb^{10}\bfx$.
		The atomic filter $\mathbf{H}_\bfa$ is real-preserving and $\mathbf{H}_\bfb$ is not.}
	\label{fig:compare-real}
\end{figure*}
\begin{figure*}[!t]
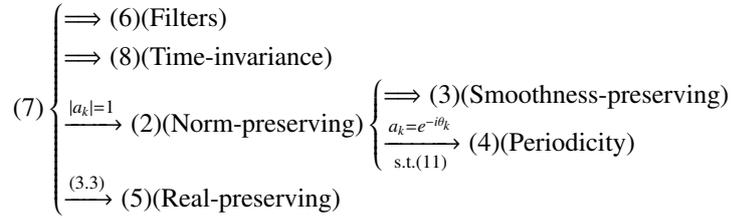

	$$(7)
	\begin{cases}
		\Longrightarrow (6)\textrm{(Filters)}\\
		\Longrightarrow (8)\textrm{(Time-invariance)}\\
		\xrightarrow{|a_k|=1}(2)\textrm{(Norm-preserving)}
		\begin{cases}
			\Longrightarrow (3) \textrm{(Smoothness-preserving)}\\
			\xrightarrow[\mathrm{s.t.} \eqref{eq:period}]{a_k=e^{-i\theta_k}}(4)\textrm{(Periodicity)}
		\end{cases}\\
		\xrightarrow{\eqref{th:Real-Shift}} (5)\textrm{(Real-preserving)}
	\end{cases}$$
	\caption{The implication relations of Properties (1)--(8). }
	\label{fig:relations}
\end{figure*}

\begin{thm}\label{th:Real-Shift}
	An atomic filter $\mathbf{H}_\mathbf{a}$ is a real matrix if and only if there exist
	a rearrangement of $\{2,\cdots,N\}$, denoted by $\{p_2,\cdots,p_N\}$, and $c_2,\cdots,c_N\in
	\bc$ with unit moduli,
	such that
	\begin{equation}\label{eq:Real-Shift-1}
		a_1\in\br,~~~~a_k=\overline{a}_{p_k},
		~~~\mathbf{u}_k=c_k\overline{\mathbf{u}}_{p_k},~~~k=2,\cdots,N.
	\end{equation}
\end{thm}
\prf
It is easy to see that $\mathbf{H}_\mathbf{a}$ is real-preserving if
and only if $\mathbf{H}_\mathbf{a}$ is a real matrix:
$\overline{\mathbf{H}}_\mathbf{a}=\mathbf{H}_\mathbf{a}$, or equivalently
$\diag(\overline{\mathbf{a}})\mathbf{U}^T\mathbf{U}
=\mathbf{U}^T\mathbf{U}\diag(\mathbf{a})$. The last equality can be rewritten as
\begin{equation}\label{eq:Real-Shift-2}
	(a_k-\overline{a}_j)\mathbf{u}_j^T\mathbf{u}_k=0,~~~~\forall\, j,k=1,\cdots,N.
\end{equation}
\par
Necessity: Assume that Eq.~\eqref{eq:Real-Shift-2} is satisfied. Since
$\mathbf{u}_1=\bfone/\sqrt{N}$, where $\bfone\in\br^N$ is an $N$-dimensional vector
of ones, then $\mathbf{u}_1^T\mathbf{u}_1=1$. By Eq.~\eqref{eq:Real-Shift-2} we deduce $a_1=\overline{a}_1$, i.e., $a_1\in\br$.
\par
For any $2\le k\le N$, since $\mathbf{u}_1^T\mathbf{u}_k=\mathbf{u}_1^*\mathbf{u}_k=0$,
if $\mathbf{u}^T_j\mathbf{u}_k=0$ holds for all $j=2,\cdots,N$,
we have that $\mathbf{U}^T\mathbf{u}_k=\mathbf{0}$, which yields an absurd conclusion
$\bfu_k=\mathbf{0}$. Thus, there exists $2\le p_k\le N$ such that
$\mathbf{u}^T_{p_k}\mathbf{u}_k\neq0$, which together with Eq.~\eqref{eq:Real-Shift-2}
implies $a_k=\overline{a}_{p_k}$. This is the second equality of Eq.~\eqref{eq:Real-Shift-1}.
\par
For $j=1,\cdots,N$,  if $a_k=\overline{a}_j$ then $a_{p_k}=a_j$,
or equivalently $j=p_k$. This shows that $a_k\neq\overline{a}_j$ if $j\neq p_k$.
Inserting it into Eq.~\eqref{eq:Real-Shift-2} we obtain that
$\mathbf{u}_j^*\overline{\mathbf{u}}_k=\overline{\mathbf{u}_j^T\mathbf{u}_k}=0$ for $j\neq p_k$, which implies that
$\overline{\bfu}_k\in\spann\{\mathbf{u}_{p_k}\}$, or equivalently,
$\overline{\bfu}_k=c_k\bfu_{p_k}$ for some $c_k\in\bc$ with unit modulus.
The third equality of Eq.~\eqref{eq:Real-Shift-1} is proved.
\par
Sufficiency: Suppose Eq.~\eqref{eq:Real-Shift-1} holds.
When $j=1$ or $k=1$, it is easy to verify that Eq.~\eqref{eq:Real-Shift-2} holds since
$a_1\in\br$ and $\mathbf{u}_1\in\br^N$.
For any $j, k=2,\cdots,N$, if $j=p_k$ then $\overline{a}_j=\overline{a}_{p_k}=a_k$, Eq.~\eqref{eq:Real-Shift-2} holds obviously;
Otherwise, there holds $j\neq p_k$, using
$\mathbf{u}_k=c_k\overline{\mathbf{u}}_{p_k}$ we obtain
$$\mathbf{u}_j^T\mathbf{u}_k=c_k\mathbf{u}_j^T\overline{\mathbf{u}}_{p_k}
=c_k\overline{\mathbf{u}_j^*\mathbf{u}_{p_k}}=0,$$
which implies Eq.~\eqref{eq:Real-Shift-2}. In summary, Eq.~\eqref{eq:Real-Shift-2} always
holds, which proves that $\mathbf{H}_\mathbf{a}$ is a real matrix.
\bbox

\par\noindent
\textbf{Remark:} If $\{\bfu_k\}_{k=1}^N$ are real-valued vectors,
it is obvious that $\{\bfu_k\}_{k=1}^N$ satisfies the last equality of
Eq.~\eqref{eq:Real-Shift-1},  i.e., $\mathbf{u}_k=\overline{\mathbf{u}}_{p_k}$
for $p_k=k$ and $c_k=1$, $k=2,\cdots,N$.


\par
Let us illustrate the result of Theorem \ref{th:Real-Shift} by experiments. Consider a ring graph,
its graph Fourier basis $\{\bfu_k\}_{k=1}^N$ is the discrete Fourier basis defined in Eq.~\eqref{eq-FT}.
It is easy to verify that $\{\bfu_k\}_{k=1}^N$ satisfy Eq.~\eqref{eq:Real-Shift-1} for $p_k=N+2-k$ and $c_k=1$, $k=2,\cdots,N$.
Thus, $\mathbf{H}_\bfa$
is real-preserving if and only if $\{a_k\}_{k=1}^N$ are distinct numbers satisfying
\begin{equation}\label{eq:real-preserving-ring-G}
	a_1\in\br~~~\mbox{and}~~~~a_k=\overline{a}_{N+2-k},~~k=2,\cdots,N.
\end{equation}
Now we set $\bfa:=\big(1,e^{i2\pi/N},\cdots,e^{i2\pi(N-1)/N}\big)^T$. It is easy to see that
Eq.~\eqref{eq:real-preserving-ring-G} holds. Next, we disorder
the components of the vector $\bfa$, denoted by $\bfb=(b_1,\cdots,b_N)^T$,
such that $\{b_k\}_{k=1}^N$ do not satisfy Eq.~\eqref{eq:real-preserving-ring-G}.
It is easy to see that $\mathbf{H}_\bfa$ and $\mathbf{H}_\bfb$ are both periodic
but $\mathbf{H}_\bfa$ is real-preserving and $\mathbf{H}_\bfb$ is not.
Fig.~\ref{fig:compare-real} plots a Gaussian signal $\bfx$ on a ring graph
and its filtered versions $\bfH_\bfa^{10}\bfx$ and $\bfH_\bfb^{10}\bfx$.
As is shown, $\bfH_\bfa^{10}\bfx$ is real-valued but $\bfH_\bfb^{10}\bfx$
is complex-valued.
\par
The relationship between the properties (1)--(8) is illustrated in Fig.~\ref{fig:relations}.


\subsection{Normal Atomic Filters}
The shift operator in the classical signal processing satisfy all the properties
(1)--(8). For atomic filters in graph signal processing, these properties may not always be satisfied
simultaneously. In fact, according to Theorem \ref{th:Real-Shift}, an atomic filter is not real-preserving
if the Fourier basis does not satisfy \eqref{eq:Real-Shift-1}. In this section, we discuss under what
conditions an atomic filter satisfies Properties (2), (4) and (5) simultaneously.
\par
Fig.~\ref{fig:relations} reveals the logical relationship of Properties (1)--(8). It shows that
both Property (2) (norm-preserving) and (5) (real-preserving) are crucial ones of atomic filters.
In this section, we shall discuss the existence and construction of such atomic filters. For
simplicity of expression, we introduce the concept of \emph{normal} atomic filter in the following definition.
\begin{de}
	An atomic filter is called normal if it is both norm-preserving and real-preserving.
\end{de}
\par
Property (5), i.e., real-preserving, is a very strong condition in graph signal processing.
If the graph Fourier basis $\{\bfu_k\}_{k=1}^N$ are real vectors, then the rearrangement
$\{p_2, \cdots, p_N\}$ of  $\{2,\cdots,N\}$ in Theorem \ref{th:Real-Shift} is exactly
$\{2,\cdots,N\}$ itself, that is, $p_k=k$ for $k=2,\cdots, N$. This fact implies that
$a_k=\overline{a}_{p_k}=\overline{a}_k$, or equivalently,  $\{a_k\}$ are
real numbers. Since there are only two different real numbers on the unit circle, by
Theorem \ref{th:isometricity} it is concluded that any real-preserving atomic filter must
not be norm-preserving if $N\ge 3$. Thus, for $N\ge 3$, to ensure the existence of normal atomic filters, the graph Fourier basis $\{\bfu_k\}_{k=1}^N$
must be selected as complex vectors. This gives us a new insight to the commonly
seen fact that the discrete Fourier basis  in the classical signal processing is a family of
complex vectors of the form \eqref{eq-FT}. For general graphs, the conditions for
the existence of normal atomic filters are given by the following theorem.
\begin{thm}\label{th:eigenvalue-multiplicity}
	Given a undirected and weighted graph, there exist normal atomic filters if and only if one of the
	following conditions is satisfied:
	\par
	(1) The Laplacian matrix $\mathbf{L}$ has at most one nonzero eigenvalue of odd multiplicity;
	\par
	(2) There exist a family of orthonormal eigenvectors
	$\{\bfu_k\}^N_{k=1}$  of $\mathbf{L}$ satisfying
	\begin{equation}\label{eq:th:eigenvalue-multiplicity}
		\bfu_1=\mathbf{1}/\sqrt{N},~~~~
		\bfu_k:=\overline{\bfu}_{N+2-k}, ~~~k=2,\cdots,\lfloor N/2\rfloor+1,
	\end{equation}
	where $\lfloor N/2\rfloor$ denotes the largest integer not exceeding $N/2$.
\end{thm}
\prf
First, we suppose $\mathbf{H}_\bfa$ is a normal atomic  filter.
Since $\mathbf{H}_\bfa$ is real-preserving,
according to Theorem \ref{th:Real-Shift},
Eq.~\eqref{eq:Real-Shift-1} holds.
Since $\bfu_k=c_k\overline{\bfu}_{p_k}=c_k\overline{c}_{p_k}\bfu_{p_{p_k}}$,
we have that $p_{p_k}=k$, which means that $p: k\mapsto p_k$ is a bijection on
$\{2,\cdots,N\}$ satisfying $p^{-1}=p$.
\par
Let $K_0:=\{k\,|\,2\le k\le N,~k=p_k\}$. It is easy to see that $a_k\in\br$ if $k\in K_0$.
Since there are only two real numbers, $1$ and $-1$, with absolute value of $1$
and $a_1\in\br$, $K_0$ contains at most one element.
Let $0=\lambda_1<\lambda_2\le\cdots\le\lambda_N$ be all the eigenvalues of
$\mathbf{L}$ satisfying $\mathbf{L}\bfu_k=\lambda_k\bfu_k,~k=1,\cdots,N$.
If $\lambda_k\not\in\{\lambda_j\,|\,j\in K_0\}$, then for any
$2\le j\le N$ with $\lambda_j=\lambda_k$ we have $j\neq p_j$.
Multiplying the both sides of $\bfu_j=c_j\overline{\bfu}_{p_j}$ by $\mathbf{L}$
we deduce that $\lambda_j=\lambda_{p_j}$. It follows that
$\bfu_j,\bfu_{p_j}\in V_{\lambda_k}:=\{\bfx\,|\,\mathbf{L}\bfx=\lambda_k\bfx\}$, the
eigenspace associated with the eigenvalue $\lambda_k$. In conclusion, there holds
$$V_{\lambda_k}=\spann\Big\{\bigcup_{j\in K_+,\,\lambda_j=\lambda_k}\{\bfu_j,\bfu_{p_j}\}~\Big\},$$
where $K_+:=\{k\,|\,2\le k\le N,~k<p_k\}$,
which shows that the multiplicity of $\lambda_k$, that equals the dimension of
$V_{\lambda_k}$, is an even number. Thus, the condition (1) is satisfied.
\par
Then, we suppose the condition (1) is satisfied. Changing the order if necessary, we assume that
all the nonzero eigenvalues $\{\lambda_k\}^N_{k=2}$ of $\mathbf{L}$ satisfy
$$\lambda_k=\lambda_{N+2-k},~~~~k=2,\cdots,\lfloor\frac{N+1}{2}\rfloor.$$
By the spectral decomposition theorem, there exist a family of orthonormal real-valued
vectors $\bm{\alpha}_1,\cdots,\bm{\alpha}_N$ with
$\bm{\alpha}_1=\mathbf{1}/\sqrt{N}$ such that
$\mathbf{L}\bm{\alpha}_k=\lambda_k\bm{\alpha}_k$ for $k=1,\cdots, N$. Define
$\bfu_1:=\bm{\alpha}_1,
~\bfu_{k}:=\frac{1}{2}(\bm{\alpha}_k+i\bm{\alpha}_{N+2-k}),
~\bfu_{N+2-k}:=\frac{1}{2}(\bm{\alpha}_k-i\bm{\alpha}_{N+k-2}),
~k=2,\cdots, \lfloor\frac{N+1}{2}\rfloor,$
and  $\bfu_{\lfloor N/2\rfloor+1}:=\bm{\alpha}_{\lfloor N/2\rfloor+1}$ if $N$ is even. It is easy to verify that
$\bfu_1,\cdots,\bfu_N$ are orthonormal eigenvectors of $\mathbf{L}$ satisfying
Eq.~\eqref{eq:th:eigenvalue-multiplicity}. Thus, the condition (2) is satisfied.
\par
Finally, we suppose the condition (2) is satisfied.  Choosing $N$ distinct complex numbers on the
unit circle of the complex plane satisfying
$$a_1=1,~~~~a_k=\overline{a}_{N+2-k},~~k=2,\cdots, \lfloor\frac{N+1}{2}\rfloor,$$
and $a_{\lfloor N/2\rfloor+1}=-1$ if $N$ is even, we know Eq.~\eqref{eq:Real-Shift-1} holds and
$|a_k|=1,~k=1,\cdots,N$. By Theorems \ref{th:isometricity} and
\ref{th:Real-Shift} we know $\mathbf{H}_\bfa$ is norm-preserving and real-preserving, that is, $\mathbf{H}_\bfa$ is a normal atomic  filter.
\bbox
\par
It is a natural question to ask what graphs meet this
condition of Theorem \ref{th:eigenvalue-multiplicity} and how to choose a vector $\bfa\in\bc^N$
such that $\bfH_\bfa$ is a normal atomic filter.
The latter will be answered in the following theorem and the first will be discussed in detail in
the next two sections.
\begin{coro}\label{coro:period-shift}
	Let $\{\bfu_k\}_{k=1}^N$ be a graph Fourier basis
	satisfying Eq.~\eqref{eq:th:eigenvalue-multiplicity}. Then for any
	$\bfa=(a_1,\cdots, a_N)^T$ with $a_1=1$,  $\mathbf{H}_\bfa$ is a normal atomic filter
	if and only if $a_k=e^{-i\theta_k},~k=1,\cdots, N$, for distinct numbers
	$\{\theta_k\}^N_{k=1}$ in $ [0,2\pi)$ satisfying
	\begin{equation}\label{eq:coro:period-shift-1}
		\theta_1=0,~~~~
		\theta_{N+2-k}+\theta_k=2\pi, ~~~k=2,\cdots, \lfloor N/2\rfloor+1.
	\end{equation}
	Furthermore, $\mathbf{H}_\mathbf{a}$ is a periodic if and only if
	Eq.~\eqref{eq:coro:period-shift-1} holds and
	\begin{equation}\label{eq:coro:period-shift-2}
		\{\theta_1,\cdots,\theta_N\}=\Big\{\frac{2\pi}{N}(k-1)\,\Big|\,k=1,\cdots,N\Big\}.
	\end{equation}
\end{coro}
\prf
From the proof of Theorem \ref{th:eigenvalue-multiplicity}, we know that $\mathbf{H}_\bfa$ is a normal atomic  filter
if and only if $\{a_k\}_{k=1}^N$ are distinct complex numbers on the unit circle of the complex plane
satisfying $a_1\in\br$ and $a_k=\overline{a}_{N+2-k},~k=2,\cdots, \lfloor N/2\rfloor+1$.
Write $a_k=e^{i\theta_k}$ with $\theta_k\in [0,2\pi)$. It is easy to see that
$a_1=1$ is equivalent to $\theta_1=0$ and
$a_k=\overline{a}_{N+2-k}$ is equivalent to
$\theta_{N+2-k}+\theta_k=2\pi$. Thus, $\mathbf{H}_\bfa$ is a normal atomic  filter if and only if
Eq.~\eqref{eq:coro:period-shift-1} holds.
\par
Furthermore, if $\mathbf{H}_\bfa$ is also periodic, then
by Proposition \ref{th:isometricity} there exist
$$\{\phi_k\,|\,k=1,\cdots,N\}=\Big\{\frac{2\pi}{N}(k-1)\,\Big|\,k=1,\cdots,N\Big\}$$
such that $a_k=e^{i\phi_k},~k=1,\cdots,N$.
Since
$\theta_k,\phi_k\in [0,2\pi)$ and $e^{i\theta_k}=e^{i\phi_k}$, we obtain that
$\theta_k=\phi_k$, which implies Eq.~\eqref{eq:coro:period-shift-2} immediately.
\bbox
\begin{figure*}[!t]
	\centerline{\includegraphics[width=0.8\textwidth]{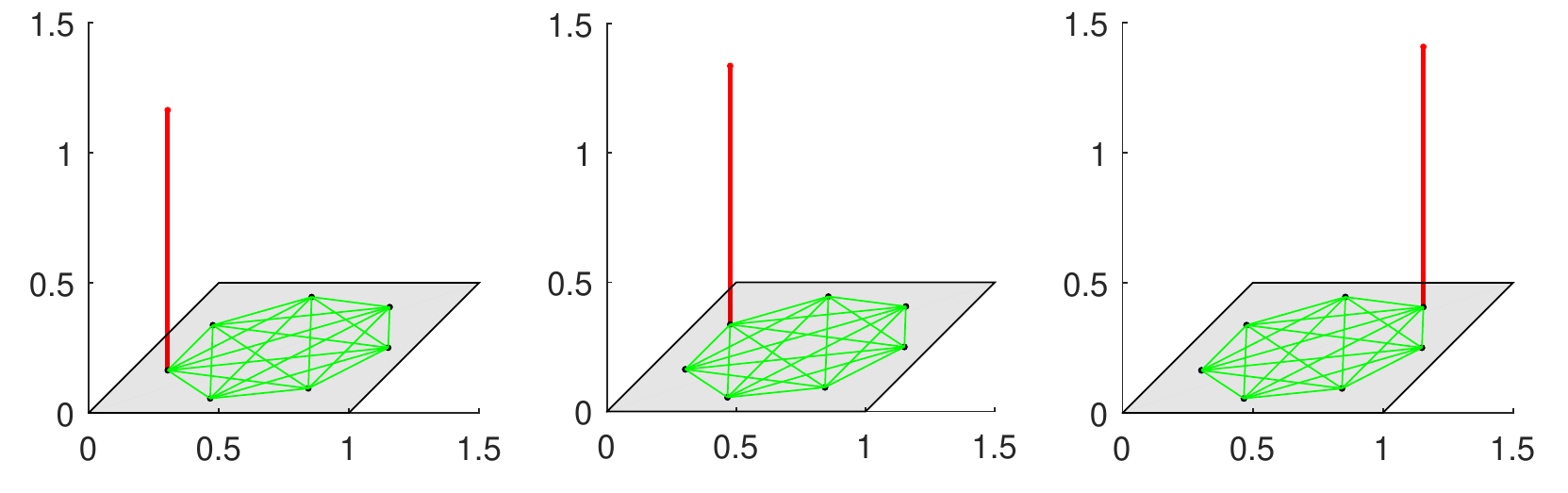}}
	\caption{\small
		Left: a pulse signal $\bfx$ defined on a fully connected graph;
		Middle: $\mathbf{H}_\bfa\bfx$; Right: $\mathbf{H}_\bfa^3\bfx$.
		The atomic filter $\mathbf{H}_\bfa$ is a periodic normal atomic  filter
		given by $\bfa=\big(1,e^{i\frac{2\pi}{N}},e^{i\frac{4\pi}{N}},\cdots,e^{i\frac{2\pi(N-1)}{N}}\big)^T$.}
	\label{fig:circulant-delta-period}
\end{figure*}
\begin{figure*}[!t]
	\centerline{\includegraphics[width=0.8\textwidth]{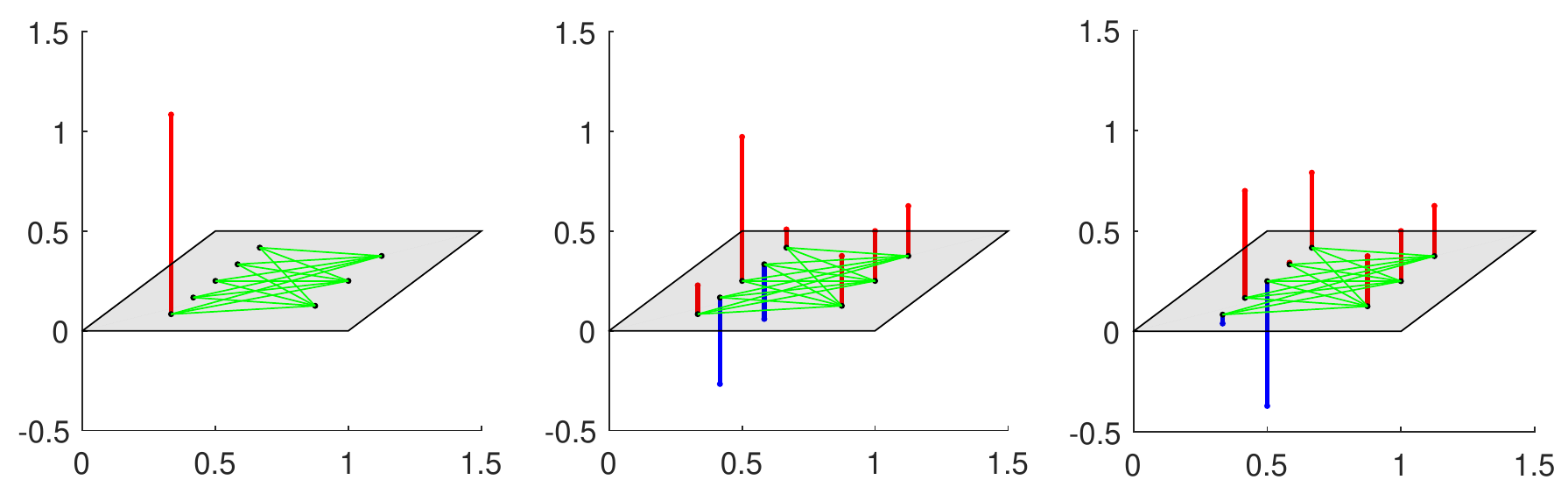}}
	\caption{\small
		Left: an impulse signal $\bfx$ defined on a complete bipartite graph
		with $p=5$ and $q=3$; Middle: $\mathbf{H}_\bfa\bfx$;
		Right: $\mathbf{H}_\bfa^3\bfx$.
		The atomic filter $\mathbf{H}_\bfa$ is a periodic normal atomic  filter
		given by $\bfa=\big(1,e^{i\frac{2\pi}{N}},e^{i\frac{4\pi}{N}},\cdots,e^{i\frac{2\pi(N-1)}{N}}\big)^T$.}
	\label{fig:bipartite-delta-period}
\end{figure*}

\section{Typical graphs with normal atomic filters}
In this section, we give some typical graphs that having normal atomic filters.
\subsection{Circulant Graphs}
\label{sec:Circulant-Graphs}
The graphs that having normal atomic filters that we want to discuss first are
the so-called circulant graphs.
An undirected weighted graph is called circulant if its adjacency matrix
$\mathbf{W}=(w_{ij})$ is a symmetric and circulant matrix,
that is, there exists a vector $\bfc=(c_0,c_1,\cdots,c_{N-1})^T$ satisfying $c_0=0$ and
$c_{N-k}=c_k$ for $k=1,\cdots,N-1$, such that
$$w_{ij}=c_{j-i\,(\mathrm{mod}~N)}\,, ~~~~i,j=1,\cdots,N.$$
The vector $\bfc\in\bc^N$ is called the generating vector of the circulant matrix $\mathbf{W}$.
It is easy to see that the generating vector $\bfc\in\bc^N$ is completely determined by its
first half components $c_1,\cdots,c_{\lfloor N/2\rfloor}$.
\par
It is easy to verify that the Laplacian matrix $\mathbf{L}$ of a circulant graph
is also a circulant matrix, whose generating vector is given by
$(\alpha, -c_1,\cdots,-c_{N-1})^T$,
where $\alpha:=\sum^{N-1}_{k=1}c_k$.
According to the spectral decomposition of circulant matrices \cite{Horn-bk2013}, we have
$\mathbf{L}=\mathbf{U}\bf{\Lambda} \mathbf{U}^{-1}$,
where $\mathbf{U}:=(\bfu_1,\cdots,\bfu_N)$ is the matrix of discrete Fourier basis defined
by Eq.~\eqref{eq-FT}.
It is easy to verify that $\{\bfu_k\}_{k=1}^N$ satisfy the condition \eqref{eq:th:eigenvalue-multiplicity},
which implies the existence of normal atomic filters.
Particularly, if $a_k = e^{-i\theta_k},~k=1,\cdots,N$,
with $\{\theta_k\}^N_{k=1}$ satisfying Eq.~\eqref{eq:coro:period-shift-1} and Eq.~\eqref{eq:coro:period-shift-2}, then by
Corollary \ref{coro:period-shift}, $\bfH_\bfa$ is also periodic. If we choose
$$\theta_k=\frac{2\pi}{N}(k-1),~~~~k=1,\cdots,N,$$
a simple calculation shows that $\bfH_\bfa$ is exactly the classical shift operator:
$(\bfH_\bfa\mathbf{x})(n)=\mathbf{x}(n-1)$.
\par
A typical example of circulant graphs is the ring graph
with adjacency matrix the circulant matrix generated by $(0,1,0,\cdots,0,1)^T$.
Another example is the fully connected graph with the adjacency matrix
the circulant matrix generated by $(0,1,\cdots,1)^T$.
For the periodic normal atomic  filter $\mathbf{H}_\bfa$ with
$\bfa=\big(1,e^{i\frac{2\pi}{N}},e^{i\frac{4\pi}{N}},\cdots,e^{i\frac{2\pi(N-1)}{N}}\big)^T$,
Fig.~\ref{fig:compare-real} in Section 2 plots $\mathbf{H}_\bfa^{10}\bfx$
for a Gaussian signal $\bfx$ defined on a ring graph,
Fig.~\ref{fig:circulant-delta-period} plots $\mathbf{H}_\bfa\bfx$
and $\mathbf{H}_\bfa^3\bfx$ for a pulse signal $\bfx$ defined on a fully connected graph.

\begin{figure*}[!t]
	\centerline{\includegraphics[width=0.8\textwidth]{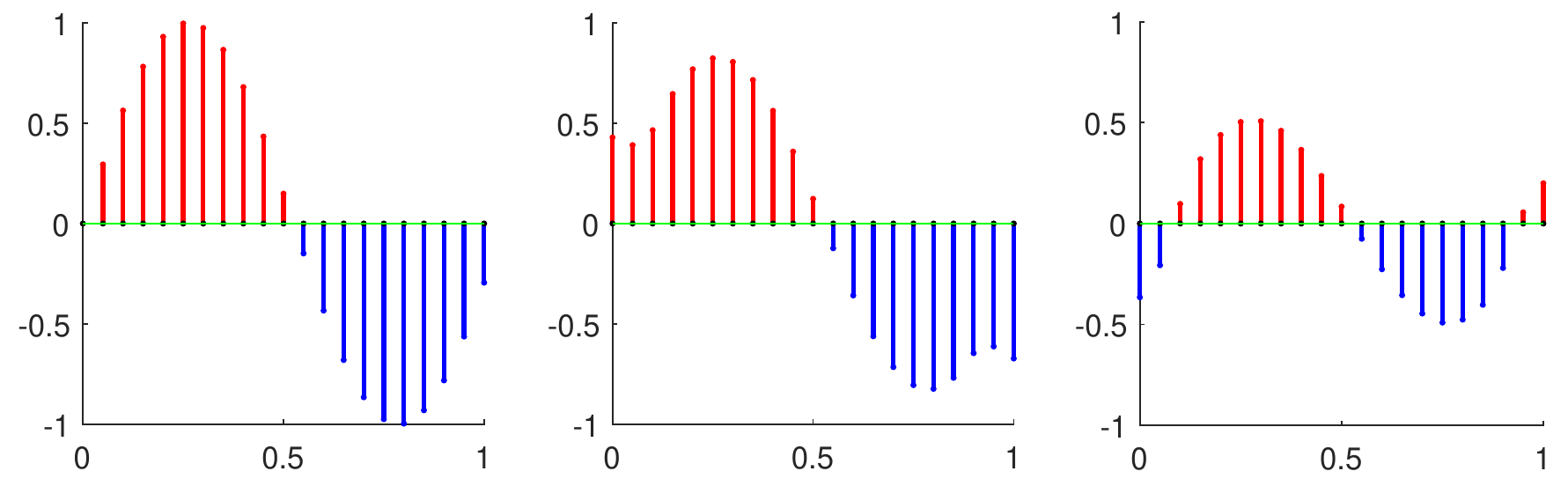}}
	\caption{\small
		Left: a sinusoidal signal $\bfx$ defined on a path graph; Middle: the real part of
		$\mathbf{H}_\bfa\bfx$; Right: the imaginary part of
		$\mathbf{H}_\bfa\bfx$.
		The atomic filter $\mathbf{H}_\bfa$ is given by $\bfa=\big(1,e^{i\frac{2\pi}{N}},e^{i\frac{4\pi}{N}},\cdots,e^{i\frac{2\pi(N-1)}{N}}\big)^T$,
		it is norm-preserving and periodic but not real-preserving.
	}
	\label{fig:path-sin-non-real}
\end{figure*}
\begin{figure*}[!t]
	\centerline{\includegraphics[width=0.8\textwidth]{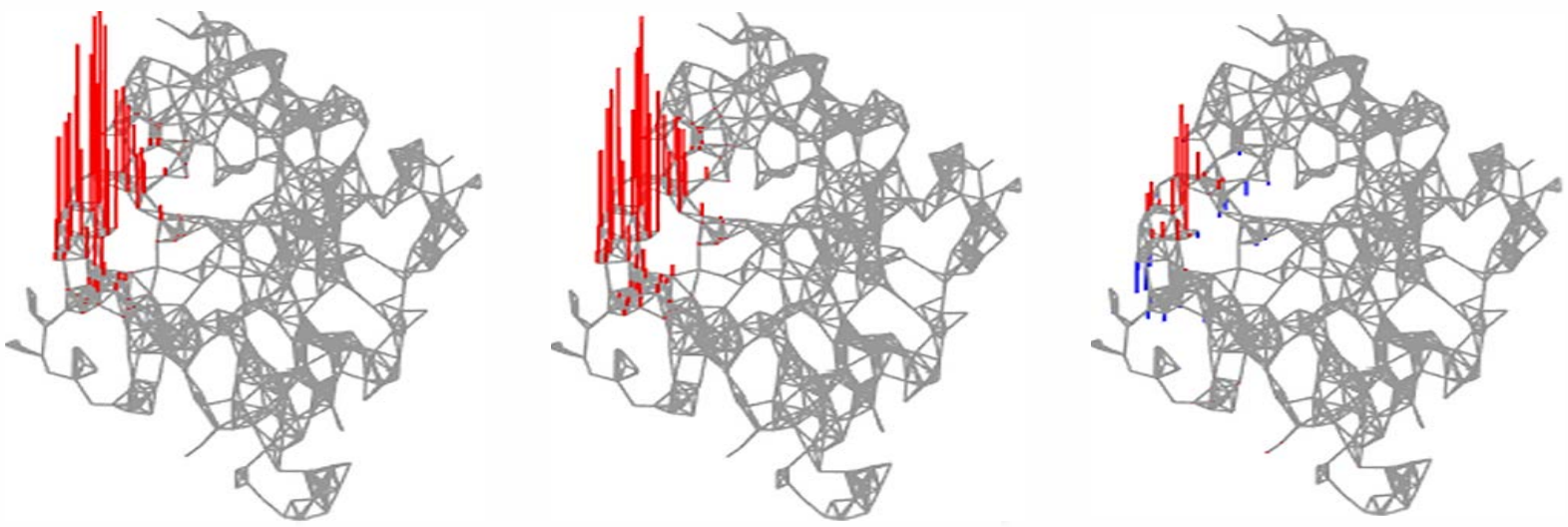}}
	\caption{\small
		Left: a Gaussian signal $\bfx$ defined on
		a sensor graph with $N=500$ vertices; Middle: the real part of
		$\mathbf{H}_\bfa\bfx$; Right: the imaginary part of
		$\mathbf{H}_\bfa\bfx$.
		The atomic filter $\mathbf{H}_\bfa$ is given by $\bfa=\big(1,e^{i\frac{2\pi}{N}},e^{i\frac{4\pi}{N}},\cdots,e^{i\frac{2\pi(N-1)}{N}}\big)^T$,
		it is norm-preserving and periodic but not real-preserving.}
	\label{fig:sensor-Gauss-non-real}
\end{figure*}

\subsection{Complete Bipartite Graphs}
A complete bipartite graph is a graph whose vertices can be partitioned into two
subsets $\mathcal{V}_1$ and $\mathcal{V}_2$ such that no edge has both endpoints
in the same subset, and every possible edge that could connect vertices in different
subsets is part of the graph. The adjacent matrix of a complete bipartite
graph with $N$ vertices has the form
$$\begin{array}{c@{\hspace{-5pt}}l}
	\mathbf{W}=\left(\begin{array}{ccc;{2pt/2pt}ccc}
		0& \cdots & 0 & 1 & \cdots & 1 \\
		\vdots& \ddots & \vdots & \vdots & \ddots &\vdots\\
		0&\cdots & 0 & 1 &\cdots & 1  \\ \hdashline [2pt/2pt]
		1& \cdots & 1 & 0 & \cdots & 0 \\
		\vdots& \ddots & \vdots & \vdots & \ddots &\vdots\\
		1&\cdots & 1 & 0 &\cdots & 0
	\end{array}\right)
	& \begin{array}{l}
		\left.\rule{0mm}{7mm}\right\}p\\
		\\\left.\rule{0mm}{7mm}\right\}q
	\end{array}\\[-5pt]
	\begin{array}{cc}
		~~~~~~~\underbrace{\rule{17mm}{0mm}}_p&
		\underbrace{\rule{17mm}{0mm}}_q\end{array} &
\end{array}
$$
where $p, q$ are respectively the cardinalities of $\mathcal{V}_1$ and $\mathcal{V}_2$, and $N=p+q$.
It can be shown that the eigenvalues of its Laplacian matrix $\mathbf{L}$ are \cite{Brouwer-bk2012}:
$$0,~\underbrace{p,~\cdots,~p}_{q-1},~\underbrace{q,~\cdots,~q}_{p-1},~N.$$
According to Theorem \ref{th:eigenvalue-multiplicity}, there exist atomic filters that are
both norm-preserving and real-preserving if and only if both $p$ and $q$ are odd numbers.
Fig.~\ref{fig:bipartite-delta-period} plots $\mathbf{H}_\bfa\bfx$ and $\mathbf{H}_\bfa^3\bfx$
for a pulse signal $\bfx$ defined
on a complete bipartite graph with $p=5$ and $q=3$,
where $\mathbf{H}_\bfa$ is a periodic normal atomic  filter given by
$\bfa=\big(1,e^{i\frac{2\pi}{N}},e^{i\frac{4\pi}{N}},\cdots,e^{i\frac{2\pi(N-1)}{N}}\big)^T$.


\subsection{Graphs without Normal Atomic Filters}
In this section, we give examples of graphs that have no normal atomic filters.

\subsubsection{Path Graphs}
A path graph is a graph in which
the first and last vertices are only connected to one adjacent vertex,
and every other vertex is only connected to two adjacent vertices.
The adjacent matrix of a path graph is given by
\begin{equation*}
	\mathbf{W}=
	\left(
	\begin{array}{cccccc}
		0 & 1 &  &  &  &  \\
		1 & 0 & 1 &  &  &  \\
		& 1 & 0 & 1 &  &  \\
		&  & \ddots & \ddots & \ddots &  \\
		&  &  & 1 & 0 & 1 \\
		&  &  &  & 1 & 0 \\
	\end{array}
	\right)
\end{equation*}
\par

The eigenvalues of the Laplacian matrix $\mathbf{L}$ of the path
graph with $N$ vertices are \cite{Brouwer-bk2012}:
$$\lambda_k=2-2\cos\frac{k\pi}{N},~~~~k=0,1,\cdots,N-1.$$
Since these $N$ eigenvalues are distinct from each other, the multiplicity of each
eigenvalue is $1$. According to Theorem \ref{th:eigenvalue-multiplicity},
there exists no normal atomic  filter on the path graph if $N\ge 3$. Nevertheless, there
exist atomic filters, which are norm-preserving and periodic but not real-preserving,
on this kind of graph.
Fig.~\ref{fig:path-sin-non-real} plots $\mathbf{H}_\bfa\bfx$ for a sinusoidal signal $\bfx$ defined on a path graph,
where the atomic filter $\mathbf{H}_\bfa$ is given by $\bfa=\big(1,e^{i\frac{2\pi}{N}},e^{i\frac{4\pi}{N}},\cdots,e^{i\frac{2\pi(N-1)}{N}}\big)^T$.
Since $\mathbf{H}_\bfa$ is norm-preserving and periodic but not real-preserving,
part of the energy of $\mathbf{H}_\bfa\bfx$ is transferred to the imaginary part.

\subsubsection{Sensor Graphs}
A sensor graph is a graph whose vertices are placed randomly in
the unit square, and edges are placed between any vertices within a fixed
radius of each other. The edge weights are assigned via a thresholded
Gaussian kernel.
Using the graph signal processing toolbox GSPBOX \cite{GSPBOX},
we find that many eigenvalues of the Laplacian matrix of the sensor graph have a multiplicity of $1$.
According to Theorem \ref{th:eigenvalue-multiplicity},
there exists no normal atomic  filter on the sensor graph.
Fig.~\ref{fig:sensor-Gauss-non-real} plots $\mathbf{H}_\bfa\bfx$ for a Gaussian signal $\bfx$ defined on
a sensor graph with $N=500$ vertices,
where the atomic filter $\mathbf{H}_\bfa$ is given by $\bfa=\big(1,e^{i\frac{2\pi}{N}},e^{i\frac{4\pi}{N}},\cdots,e^{i\frac{2\pi(N-1)}{N}}\big)^T$.
Since $\mathbf{H}_\bfa$ is norm-preserving and periodic but not real-preserving,
part of the energy of $\mathbf{H}_\bfa\bfx$ is transferred to the imaginary part.

\section{Relations to Existing Works in the Literature}
In this section, we make a discussion of the relations between our work and those
in the literature.
\par
In \cite{Shuman-2016}, the graph shift $\mathbf{T}_k$ is defined via generalized convolution with a pulse signal located at vertex $v_k$.
It is actually a graph filter, that is, for $k=1,\cdots,N$,
\begin{equation*}
	\mathbf{T}_k=\bfU\diag(\bfa)\bfU^{-1},\\
	\mbox{where}~~\bfa=\sqrt{N}\Big(\overline{\bfu_1(k)},\cdots,\overline{\bfu_N(k)}\Big)^T.
\end{equation*}
where $\bfU$ is the matrix of graph Fourier basis.
For each $k$, if the $k$-th row of $\bfU$ has distinct components,
$\mathbf{T}_k$ is an atomic filter.
Generally, it is not norm-preserving since
$|\bfu_1(k)|=\cdots=|\bfu_N(k)|=1/\sqrt{N}$ does not hold.
Accordingly, it is not a normal atomic  filter.
Moreover, the composability $\mathbf{T}_{k+l}=\mathbf{T}_k\mathbf{T}_l$
usually does not hold.
\par
In \cite{Sandryhaila-2013}, the authors regarded the adjacency matrix $\mathbf{W}$
of a graph as the graph shift.
In the special case of directed ring graphs, it is easy to show that
$\mathbf{W}$ is consistent with the shift operator in classical signal processing.
For directed graphs, it is incomparable with the proposed atomic filter or
normal atomic  filter since the graphs we consider here are undirected. For undirected graphs,
we claim that
$\mathbf{W}$ is a graph filter if and only if the graph is regular.
\par
In fact, if $\mathbf{W}=\bfH_\bfa:=\bfU\diag(\bfa)\bfU^{-1}$ for some
$\bfa\in\bc^N$, then
$$\mathbf{D}=\mathbf{L}+\mathbf{W}=\bfU(\bm{\Lambda}+\diag(\bfa))\bfU^{-1}.$$
Let $\mathbf{D}=\diag(\mathbf{d})$, where $\mathbf{d}:=(d_1,\cdots,d_N)^T$.
Let $\lambda_1,\cdots,\lambda_N$ be all the eigenvalues of the Laplacian matrix $\bfL$.
Then $(d_i-\lambda_j-a_j)u_{ij}=0$ for $i,j=1,\cdots, N$, where $(u_{ij})=\bfU$. Since
$u_{i1}=1/\sqrt{N},~i=1,\cdots,N$, we have that $d_i=\lambda_1+a_1=a_1$
for $i=1,\cdots,N$, which implies that the graph is $a_1$-regular.
Conversely, if the graph is $d$-regular, then
$\mathbf{W}=\mathbf{D}-\mathbf{L}=\bfU(d\,\mathbf{I}_N-\bm{\Lambda})\bfU^{-1}$ is
a filter with the frequency response $\bfa:=(d-\lambda_1,d-\lambda_2,\cdots,d-\lambda_N)^T$.
\par
Furthermore, if the graph is $d$-regular, it is easy to see that $\mathbf{W}$ is
an atomic filter if and only if the eigenvalues $\lambda_1,\cdots,\lambda_N$ of $\mathbf{L}$ are different
from each another. In this case, it is obviously real-preserving.
Meanwhile, since the frequency response $\bfa\in\bc^N$ is a real vector,
$\mathbf{W}$ is not norm-preserving if $N\ge 3$, thus it is not a normal atomic  filter.
\par
In \cite{Girault-2015,Girault-bk2015,Girault-2016}, the authors defined the graph shift by
\begin{align*}
	&\mathbf{T}_{\mathcal{G}}:=\bfU\diag(\bfa)\bfU^{-1},\\
	&\mbox{where}~~\bfa=\big(e^{-i\pi\sqrt{\lambda_1/\rho_\mathcal{G}}},\cdots,e^{-i\pi\sqrt{\lambda_{N}/\rho_\mathcal{G}}}\big)^T,
\end{align*}
where $\lambda_1\le\cdots\le\lambda_{N}$ are
the eigenvalues of the graph Laplacian matrix $\mathbf{L}$, and
$\rho_\mathcal{G}$ is an upper bound on the eigenvalues of $\mathbf{L}$.
If $\lambda_1,\cdots,\lambda_N$ are different from each another,
then $\mathbf{T}_{\mathcal{G}}$ is a norm-preserving atomic filter.
In general it is neither periodic nor real-preserving.
\par
In \cite{Gavili-2017}, the authors defined a set of graph shifts by
\begin{equation*}
	\mathbf{T}_{\bf{\phi}}:=\bfU\diag(\bfa)\bfU^{-1},~~~~\mbox{where}~~\bfa=\big(e^{i\phi_1},\cdots,e^{i\phi_N}\big)^T,
\end{equation*}
where $\phi_1,\cdots,\phi_N\in[0,2\pi)$ are distinct numbers. It is a norm-preserving atomic filter.
By Proposition \ref{th:isometricity}, if
$\{\phi_k\,|\,k=1,\cdots,N\}=\big\{\frac{2\pi}{N}(k-1)\,\big|\,k=1,\cdots,N\big\}$, then
it is periodic. The real-preserving property is not taken into account in \cite{Gavili-2017}.
\par
In \cite{Chen-2018},
the authors studied the shift-enabled condition that the characteristic and minimum
polynomials of the matrix $\mathbf{H}$ are identical.
They prove that $\mathbf{H}$ is shift-enabled if and only if every
linear shift-invariant operator is a polynomial of $\mathbf{H}$.
Since an atomic filter $\bfH_\bfa$ has distinct eigenvalues $a_1,\cdots,a_N$,
it satisfies the shift-enabled condition.
However, it is easy to understand that a shift-enabled matrix $\mathbf{H}$ is not
necessarily an atomic filter since it may even not be a graph filter.
\par
In \cite{Pesenson-2010b}, by using the Schr\"{o}dinger's semigroup of operators generated by the
graph Laplacian $L$, I. Z. Pesenson \textit{et al.} defined a family of shift operators
$$T_h := e^{ihL}=\mathbf{U}\diag(\mathbf{a}_h)\mathbf{U}^{-1},$$
where $\mathbf{a}_h:=(e^{i\lambda_1h},...,e^{i\lambda_Nh})^T$ and $h\in\br$.
Inspired by the idea of this definition, a similar definition was
proposed in \cite{huang2021approximation} as
$$\tilde{T}_h := e^{ih\sqrt{L}}=\mathbf{U}\diag(\tilde{\mathbf{a}}_h)\mathbf{U}^{-1},$$
where $\tilde{\mathbf{a}}_h:=(e^{ih\sqrt{\lambda_1}},...,e^{ih\sqrt{\lambda_N}})^T$ and $h\in\br$.
It is easy to see that both $T_h$ and $\tilde{T}_h$ are atomic filter if and only if $\mathbf{L}$
has distinct eigenvalues. If they are atomic filters, they must be norm-preserving. However,
whether they are real-preserving depends on whether they satisfy \eqref{eq:Real-Shift-1}.

\section{Windowed Fourier Time-frequency Atom for Graph Signals}
The windowed Fourier transform is a powerful tool in time-frequency analysis in
the classical signal processing. Its windowed Fourier time-frequency atom is defined
as the product of the Fourier basis function and the shift of a window function:
$$g_{b,\xi}(x):=g(x-b)e^{i\xi x},$$
where the window function $g$ is usually a compactly supported function and the shift
$b$ controls the center of the locality. It is obviously seen that the shift operator
play a crucial role. In order to extend the theory to graph signal processing, we
use normal atomic filters as an alternative to shift operator and define the
windowed Fourier time-frequency atom of the graph setting as follows:
$$\mathbf{g}_{\bfa, k}:=(\mathbf{H}_{\bfa}\mathbf{g})\odot\bfu_k,~~~k=1,\cdots,N,$$
where $\mathbf{g}\in\bc^N$ is a graph signal (window function),
$\mathbf{H}_{\bfa}$ is an atomic filter, `$\odot$' represents the Hadamard product
of two vectors.
In the following theorem, we prove that under proper conditions
$\{\mathbf{g}_{\bfa_j,k}\}_{1\le j\le J,\, 1\le k\le N}$
constitute a frame for graph signals,  and any graph signal $\mathbf{f}\in\bc^N$ can be completely
reconstructed from $\{\langle\mathbf{f},\mathbf{g}_{\bfa_j,k}\rangle\}_{1\le j\le J,\, 1\le k\le N}$.
It extends the results in \cite{Shuman-2016}.
\begin{thm}\label{th:reconstruction}
	Let $\{\bfu_k\}_{k=1}^N$ be a graph Fourier basis,
	$\mathbf{g}\in\bc^N$ and $\mathbf{A}:=(\mathbf{a}_1,\cdots,\mathbf{a}_J)
	\in\bc^{N\times J}$.
	If the row vectors of $\mathbf{A}$ are orthonormal and
	\begin{equation}\label{eq:th:reconstruction-Cn}
		C_n:=\sum_{l=1}^N|\hat{\mathbf{g}}(l) |^2|\mathbf{u}_l(n)|^2>0,
		~~~~n=1,\cdots, N,
	\end{equation}
	then for any graph signal $\mathbf{f}\in\bc^N$, the following reconstruction formula holds:
	\begin{equation}\label{eq:th:reconstruction}
		\mathbf{f}(n)
		=\frac{1}{C_n}\sum_{j=1}^J\sum_{k=1}^N \langle\mathbf{f},\mathbf{g}_{\bfa_j,k}\rangle\,\mathbf{g}_{\bfa_j,k}(n),~~~~n=1,\cdots,N,
	\end{equation}
	and
	\begin{equation}\label{th:frame-inequality}
		\alpha\|\mathbf{f}\|_2^2
		\le \sum_{j=1}^J\sum_{k=1}^N|\langle\mathbf{f},\mathbf{g}_{\bfa_j,k}\rangle|^2
		\le \beta\|\mathbf{f}\|_2^2\,,
	\end{equation}
	where $\alpha:=\min_{1\le n\le N}C_n,~\beta:=\max_{1\le n\le N}C_n$.
\end{thm}
\prf
Since the row vectors of $\mathbf{A}$ are orthonormal, i.e.
$\sum_{j=1}^J\overline{\bfa_j(l)}\,\bfa_j(l')=\delta_{l,\,l'}$ holds,
we have that
\begin{align*}
	&\sum_{j=1}^J\sum_{k=1}^N \langle\mathbf{f},\mathbf{g}_{\bfa_j,k}\rangle\,\mathbf{g}_{\bfa_j,k}(n) \\
	&=\sum_{j=1}^J\sum_{k=1}^N \Big(\sum_{m=1}^N\mathbf{f}(m)\overline{\mathbf{u}_k(m)}
	\sum_{l=1}^N\overline{\bfa_j(l)}\overline{\hat{\mathbf{g}}(l)}
	\overline{\mathbf{u}_l (m)}\Big)\\
	&~~\cdot\Big(\mathbf{u}_k(n)\sum_{l'=1}^N\bfa_j(l')\hat{\mathbf{g}}(l')
	\mathbf{u}_{l'}(n)\Big) \\
	&=\sum_{m=1}^N\mathbf{f}(m)\sum_{l=1}^N\sum_{l'=1}^N
	\overline{\hat{\mathbf{g}}(l)}\,
	\hat{\mathbf{g}}(l')\,\overline{\mathbf{u}_l (m)}\,\mathbf{u}_{l'}(n)\\
	&~~\cdot\sum_{j=1}^J\overline{\bfa_j(l)}\,\bfa_j(l')\sum_{k=1}^N\overline{\mathbf{u}_k(m)}\,\mathbf{u}_k(n) \\
	&=\sum_{m=1}^N\mathbf{f}(m)\sum_{\ell =1}^N\sum_{l'=1}^N\overline{\hat{\mathbf{g}}(l)} \,\hat{\mathbf{g}}(l')
	\,\overline{\mathbf{u}_l (m)}\,\mathbf{u}_{l'}(n)\,\delta_{l,\,l'}\,\delta_{m,n} \\
	&=\mathbf{f}(n)\sum_{l =1}^N|\hat{\mathbf{g}}(l) |^2|\mathbf{u}_l (n)|^2\\
	&=C_n\mathbf{f}(n),
\end{align*}
which yields Eq.~\eqref{eq:th:reconstruction} immediately.
Multiplying the both sides of the above equality by $\overline{\mathbf{f}(n)}$ and
summing them over $n$ from $1$ to $N$, we obtain that
$$\sum_{j=1}^J\sum_{k=1}^N |\langle\mathbf{f},\mathbf{g}_{\bfa_j,k}\rangle|^2
=\sum^N_{n=1}C_n|\mathbf{f}(n)|^2.$$
Using $\alpha\le C_n\le\beta$ in the above equality, we deduce Eq.~\eqref{th:frame-inequality}.
\bbox
\par\noindent
\textbf{Remark:}
When $C_n=C>0$ for all $n=1,\cdots,N$, $\{\mathbf{g}_{\bfa_j,k}\}_{1\le j\le J,\, 1\le k\le N}$
constitute a tight frame with frame bound $C$ and the reconstruction Eq.~\eqref{eq:th:reconstruction} can be rewritten as
\begin{equation*}
	\mathbf{f}
	=\frac{1}{C}\sum_{j=1}^J\sum_{k=1}^N \langle\mathbf{f},\mathbf{g}_{\bfa_j,k}\rangle\,\mathbf{g}_{\bfa_j,k}.
\end{equation*}
A special example of the case is when the graph is circulant. In this case,
according to Eq.~\eqref{eq-FT} we have that $|\bfu_\ell(n)|=1/\sqrt{N}$.
Thus
\begin{equation}
	C_n=\frac{1}{N}\sum_{\ell =1}^N|\hat{\mathbf{g}}(l) |^2
	=\frac{1}{N}\|\hat{\mathbf{g}}\|_2^2
	=\frac{1}{N}\|\mathbf{g}\|_2^2,
	~~~~n=1,\cdots, N.
\end{equation}
\par
Let us consider the following special family of time-frequency atoms:
\begin{equation}\label{eq:sec5-special-atoms}
	\mathbf{g}_{j, k}
	:=\frac{1}{\sqrt{N}}(\mathbf{H}_{\bfa}^{j-1}\mathbf{g})\odot\bfu_k,
	~~~j, k=1,\cdots,N,
\end{equation}
where $\bfa:=(a_1,\cdots, a_N)^T\in\bc^N$.
It is easy to see that
$$\frac{1}{\sqrt{N}}\mathbf{H}_{\bfa}^{j-1}
=\mathbf{H}_{\bfa_j},$$
where
$$\bfa_j:=\frac{1}{\sqrt{N}}(a_1^{j-1},\cdots, a_N^{j-1})^T,~~~~j=1,\cdots,N.$$
A simple calculation shows that
\begin{equation}\label{eq:sec5-matrix-A}
	\mathbf{A}:=(\bfa_1,\cdots,\bfa_N)
	=\frac{1}{\sqrt{N}}
	\begin{pmatrix}
		1 & a_1 & a_1^2 & \cdots & a_1^{N-1} \\
		1 & a_2 & a_2^2 & \cdots & a_2^{N-1} \\
		\vdots & \vdots & \vdots & \ddots & \vdots \\
		1 & a_N & a_N^2 & \cdots & a_N^{N-1}
	\end{pmatrix}.
\end{equation}
\begin{lem}\label{lem:sec5-circulant-matrix}
	For distinct numbers $a_1,\cdots,a_N\in\bc$,  the matrix
	$\mathbf{A}$ defined by Eq.~\eqref{eq:sec5-matrix-A}
	is a unitary matrix if and only if there exist a permutation matrix $\mathbf{P}$
	and a diagonal matrix $\mathbf{C}:=\diag(1,c,\cdots,c^{N-1})$ with $|c|=1$ such that
	$\mathbf{A}=\mathbf{P}\bfU \mathbf{C}$, where
	$\bfU$ is the matrix of discrete Fourier basis defined in Eq.~\eqref{eq-FT}.
\end{lem}
\prf
The sufficiency holds obviously since $\mathbf{P}\mathbf{P}^*=\bfU\bfU^*=\mathbf{C}\mathbf{C}^*=\mathbf{I}_N$.
To prove the necessity we assume that
$\mathbf{A}$ is a unitary matrix. For  $k=1,2,\cdots,N$, we have that
$1+|a_k|^2+|a_k|^4+\cdots|a_k|^{2(N-1)}=N$,
which implies that  $|a_k|=1$. Write $a_k=e^{i\omega_k}$.
Multiplying a permutation matrix $\mathbf{Q}$ on the left of $\mathbf{A}$
we can assume that $0\le\omega_1<\omega_2<\cdots<\omega_N<2\pi$.
\par
For $k=2,\cdots,N$, we have $1+\overline{a}_1a_k+\overline{a}_1^2a_k^2+
\cdots+\overline{a}_1^{N-1}a_k^{N-1}=0$, i.e.,
$$1+e^{i(\omega_k-\omega_1)}+e^{i2(\omega_k-\omega_1)}+e^{i(N-1)(\omega_k-\omega_1)}=0.$$
It follows that $e^{iN(\omega_k-\omega_1)}=1$ and consequently
$\frac{N}{2\pi}(\omega_k-\omega_1)\in\bz$.
Since
$$\Big\{\frac{N}{2\pi}(\omega_k-\omega_1)\,\Big|\,k=1,\cdots,N\Big\}\subset[0,N),$$
and there are exactly $N$ integers in $[0,N)$, we deduce that
$\frac{N}{2\pi}(\omega_k-\omega_1)=k-1$ for $k=1,\cdots,N$. Hence
$$a_k=c\,\omega^{k-1},~~k=1,\cdots,N,~~~~\mbox{where}~~
\omega:=e^{i2\pi/N},~~c:=e^{i\omega_1},$$
which is equivalent to $\mathbf{Q}\mathbf{A}=\bfU\mathbf{C}$, where $\mathbf{C}:=\diag(1,c,\cdots,c^{N-1})$. Then we obtain
$\mathbf{A}=\mathbf{P}\bfU\mathbf{C}$, where $\mathbf{P}:=\mathbf{Q}^T$ is a permutation matrix. The necessity is proved.
\bbox
\par
Lemma \ref{lem:sec5-circulant-matrix} shows that the matrix $\mathbf{A}$ defined in Eq.~\eqref{eq:sec5-matrix-A} satisfies the condition
of Theorem \ref{th:reconstruction} if and only if $\mathbf{A}$ (possibly after row permutation) is equivalent
to the classical discrete Fourier basis matrix up to a permutation.
Based on this fact we have the following corollary.
\begin{coro}
	For $\{\bfu_k\}_{k=1}^N, \mathbf{g}, C_n, \alpha,\beta$ stated in Theorem \ref{th:reconstruction}. If $C_n>0$ for $n=1,\cdots, N$, then
	for $\bfa:=(1, \omega,\cdots, \omega^{N-1})^T$ with $\omega:=e^{i2\pi/N}$,
	the time-frequency atoms
	$$\mathbf{g}_{j,k}:=\frac{1}{\sqrt{N}}(\mathbf{H}_{\bfa}^{j-1}\mathbf{g})\odot\bfu_k,
	~~~j, k=1,\cdots,N,$$
	constitute a frame of $\bc^N$ with bounds $\alpha$ and $\beta$, that is,
	$$\alpha\|\mathbf{f}\|_2^2
	\le \sum_{j=1}^J\sum_{k=1}^N
	|\langle\mathbf{f},\mathbf{g}_{j,k}\rangle|^2
	\le \beta\|\mathbf{f}\|_2^2\,,~~~~\forall\, \mathbf{f}\in\bc^N,$$
	and the reconstruction formula
	$$\mathbf{f}(n)
	=\frac{1}{C_n}\sum_{j=1}^J\sum_{k=1}^N
	\langle\mathbf{f},\mathbf{g}_{j,k}\rangle
	\,\mathbf{g}_{j,k}(n),~~~~n=1,\cdots,N.$$
	holds for any $\mathbf{f}\in\bc^N$.
\end{coro}

\section{Conclusion}
The definition of shift operation for signals defined on graphs is a fundamental problem
since the vertex set of a graph is usually not a vector space.
This paper provides a system research on this topic. A weak form, called atomic filter, is
presented. The contributions of this paper are summarized as follows:
\begin{itemize}
	\item
	Based on the action of shift operator in the classical signal processing, we introduce the concept of
	atom filters. The basic properties, including the norm-preserving, smooth-preserving, periodic,
	and real-preserving are studied.
	\item
	The property of real-preserving holds naturally in the classical signal processing, but no
	report on the researches on this topic. This paper shows that real-preserving may not
	always hold for any graph. The sufficient and necessary conditions are presented.
	\item
	The concept of normal atomic filters is introduced.
	Typical examples of graphs that have or have not normal atomic filters are given.
	\item
	Finally, as an application, atomic filters are utilized to construct time-frequency atoms
	which constitute a frame of the graph signal space.
\end{itemize}

\section*{Acknowledgement}
This work is supported by National Natural Science Foundation of China
(Grant Nos. 12171488, 11501377) and Guangdong Province Key Laboratory of Computational Science
at the Sun Yat-sen University (2020B1212060032)




  \bibliographystyle{elsarticle-num-names}
  \bibliography{UTF8-refs}






\end{document}